\documentclass[conference]{IEEEtran}

\usepackage{cite}
\usepackage{amsmath,amssymb,amsfonts}
\usepackage{amsthm}
\usepackage{algorithm}
\usepackage{graphicx}
\usepackage{algpseudocode}
\usepackage{textcomp}
\usepackage{xcolor}
\usepackage{comment}
\usepackage{stfloats}

\usepackage[margin=0.8in, top=0.8in,bottom=1.1in, columnsep=0.24in]{geometry}
\renewcommand{\baselinestretch}{0.882}\small\normalsize

\DeclareMathOperator*{\argmin}{argmin}

\DeclareMathOperator{\size}{size}
\DeclareMathOperator{\index1}{index}

\def\BibTeX{{\rm B\kern-.05em{\sc i\kern-.025em b}\kern-.08em
    T\kern-.1667em\lower.7ex\hbox{E}\kern-.125emX}}
\begin{document}
\title{Near-Pilotless Single Carrier Communications Using Matrix Decomposition\\

}


\author{\IEEEauthorblockN{K. Sai Praneeth\IEEEauthorrefmark{1},
P. Aswathylakshmi \IEEEauthorrefmark{2} and,
Radha Krishna Ganti\IEEEauthorrefmark{5}}
\IEEEauthorblockA{Department of Electrical Engineering, Indian Institute of Technology
Madras, Chennai-600036, India.\\
Email: \IEEEauthorrefmark{1}praneethk@smail.iitm.ac.in,
        \IEEEauthorrefmark{2}aswathylakshmi@ee.iitm.ac.in,
        \IEEEauthorrefmark{5}rganti@ee.iitm.ac.in
}
}

\maketitle

\begin{abstract}

Single Input-Multiple Output (SIMO) systems are key enablers of high data rates in the next generation wireless communications. However in SIMO systems, channel estimation and equalization are challenging particularly in the presence of rapidly changing channels. The high pilot overhead required for channel estimation can reduce the system throughput for large antenna configuration. In this paper, we provide an iterative matrix decomposition algorithm for near-pilotless or blind decoding of SIMO signals, in a single carrier system with frequency domain equalization. This novel approach replaces the standard equalization and estimates both the transmitted data and the channel without the knowledge of any prior distributions, by making use of only one pilot. Our simulations demonstrate improved performance, in terms of error rates, compared to the more widely used pilot-based Maximal Ratio Combining (MRC) method. 
\end{abstract}


\section{Introduction}
Single Carrier (SC) systems are extensively used in satellite communications. While commercial cellular communication systems have transitioned from single carrier to Orthogonal Frequency Division Multiplexing (OFDM) systems, SC systems continue to present several advantages over OFDM, including reduced Peak Average Power Ratio (PAPR), enhanced sensitivity to Carrier Frequency Offset (CFO), improved tolerance to amplifier nonlinearities, and a simplified design for both transmitters and receivers.

SIMO systems using large antenna arrays facilitate both multiplexing and receiver diversity. An efficient implementation of a large SIMO system requires accurate channel estimation and equalization \cite{995852}. In a conventional single carrier system, the channel is estimated by transmitting a training pilot sequence (utilising orthogonal resources for each antenna). Assuming a relatively stable channel, the same set of channel estimates is used for equalization in the subsequent transmissions. As the number of antennas increases, these pilot sequences become longer (to preserve orthogonality across the antennas and mitigate interference) resulting in a reduction in the overall system throughput. 

Single carrier systems with time domain equalization use tapped delay line filtering to eliminate Inter Block Interference (IBI). In case of wide-band channels, the length of these time domain filters increases linearly with the length of the channel impulse response. Frequency Division Equalization (FDE) is more efficient in wide-band channels. A comparative performance analysis of OFDM and single-carrier FDE is presented in  \cite{Acolatse2010DesignOS} for frequency-selective channels. In \cite{1369274} \cite{935746} and \cite{7485388}, semi-blind, training sequence based channel estimation methods for single carrier communications were proposed, which inherently use pilots or training sequences for channel estimation. 
Similarly, \cite{4773788} uses a precoding sequence to minimize the noise effects in the covariance matrix estimation, but this is applied for a SISO system. 
Also, \cite{zhang2017blindsignaldetectionmassive} presents a blind detection algorithm using approximate message passing that estimates signal and channel simultaneously, but it requires prior distributions on channel and data.
This work extends our previous research on pilotless uplink for OFDM-based massive MIMO systems \cite{Aswathylakshmi_2023} to single-carrier systems. In this paper, we present a novel technique for single-carrier systems using frequency domain equalization that is nearly pilotless (irrespective of the number of antennas). The proposed iterative algorithm uses alternating minimization to minimize between channel estimation and maximal ratio combining (MRC) for equalization. Our approach offers several advantages over existing work:
\begin{enumerate}
    \item Estimates of transmitted data and channel are calculated simultaneously using an iterative method. So, if the coherence time is large enough, the same estimates of channel can be used for decoding the next transmission.
    \item The algorithm requires a single pilot to provide an estimate of the data and channel for a massive SIMO system. This reduces the overhead, in contrast to other systems that rely on optimizing pilots \cite{tomasi2016pilot}.
    \item The proposed algorithm is scalable for multiple transmitter antennas (MIMO) which would be the future direction of this work.
    \item Unlike \cite{yuan2024integratednearfieldsensing}, which employs some prior distribution on the channel and is confined to Line Of Sight (LOS), our proposed method operates with frequency-selective channels without assuming any prior distribution on the channel.
\end{enumerate}
\textbf{Notations} : All upper case variables are matrices, whereas the lower case variables are vectors. $(.)^*$ indicates conjugate, $(.)^T$ indicates transpose, and $(.)^H$ denotes complex conjugate transpose or hermitian, $(.)_{F}$ is the Frobenius norm. $\dagger$ denotes moore penrose inverse or pseudo-inverse. $\circledast$ represents circular convolution. $(.)^{-1}$ indicates inverse. $\odot$ denotes element wise multiplication.

\section{System Model}
In this paper, we assume a single-antenna transmitter (User) and a multi-antenna receiver (Base station) with $N_r$ antennas. The channel between the transmitter and the $i$-th receive antenna is assumed to be an $L$-tap (frequency selective) channel given by $\{ {h_i}[0], {h_i}[1],\hdots, {h_i}[L-1]\}$. To aid in frequency domain equalization, we assume that the transmitter pads a cyclic prefix (CP) of length $L$.
More precisely, we assume that the transmitter sends the time domain data sequence (as a frame), ${{{x}[n]}}=\{x[P-L],\hdots,x[P-1],x[0],x[1],x[2],\hdots,x[P-1]\}$, where all the symbols are drawn from $M$-QAM constellation. The first set of symbols $\{x[P-L],\hdots,x[P-1]\}$ corresponds to the CP portion of the signal.
We also assume that $x[0]$ in every frame is a pilot symbol, i.e., the symbol is known at both the transmitter and the receiver. This pilot is used to recover the correct phase of the transmitted signal post equalization at the receiver. The received signal at antenna $i$, can be modeled as 
 \begin{equation}
    y_i[n] = x[n] \circledast {h_i}[n] +  w_i[n],
     \label{eq:1}
\end{equation}
where $w_i[n]$ is the additive white circularly symmetric complex gaussian noise (AWGN). 

Given the $P$ transmitted symbols of the user (after cyclic prefix removal) and expressing the circular convolution operation in \eqref{eq:1} using a circulant matrix obtained by appropriate zero-padding of the channel vector, the received signal at the $i^{th}$ receive antenna can be represented as:

 \begin{equation}
    y_i[n]_{(P\times 1)} = C_{x{(P\times P)}} 
    \begin{bmatrix}
        {h_i[n]} \\
        {0_{(P-L)}}
    \end{bmatrix}_{(P\times 1)}
     +  w_i[n]_{(P\times 1)},
     \label{eq:circ_matrix}
\end{equation}
where $ {C_x}$ represents the $P \times P$ column shifted circulant matrix of the transmitted data i.e, each column of the matrix $C_x$ is a circular shift by $n$ of the sequence $x[n], n=0,\hdots,P-1$ and ${h_i}[n]$ is the $L\times 1$ channel response vector and is padded with $P-L$ zeros. By arranging the received signal across $N_r$ antennas in matrix form, we get
\begin{equation} \label{eq:2}
    { {Y}}_{(P\times N_r)} = 
    {C_x}_{(P\times P)}
    \begin{bmatrix}
        {h}_{L\times N_r} \\
        {0_{(P-L)\times N_r}}
    \end{bmatrix}_{(P\times N_r)}  
    {+ {W}}_{(P\times N_r)},
\end{equation} 
where, $h_{L\times N_r}$ is the time domain channel across $N_r$ receiver antennas. Let, $H =  \begin{bmatrix}
        {h}_{L\times N_r} \\
        {0_{(P-L)\times N_r}}
    \end{bmatrix} $. 
Thus,
\begin{equation} \label{eq:7}
    {Y = C_x H + W},
\end{equation}
 Since ${C_x}$ is a square circulant matrix, it can be decomposed using the Discrete Fourier Transform (DFT) matrices \cite{golub13} as
\begin{equation} \label{eq:5}
    {C_x = F^H \Lambda_x F}.
    \end{equation}
where, $F$ is a $P\times P$ DFT matrix. Using \eqref{eq:5} we have, 
\begin{equation} \label{eq:7}
    {Y = F^H \Lambda_x F H + W},
\end{equation}

\begin{figure*}[ht!]
\centerline{\includegraphics[scale=0.45]{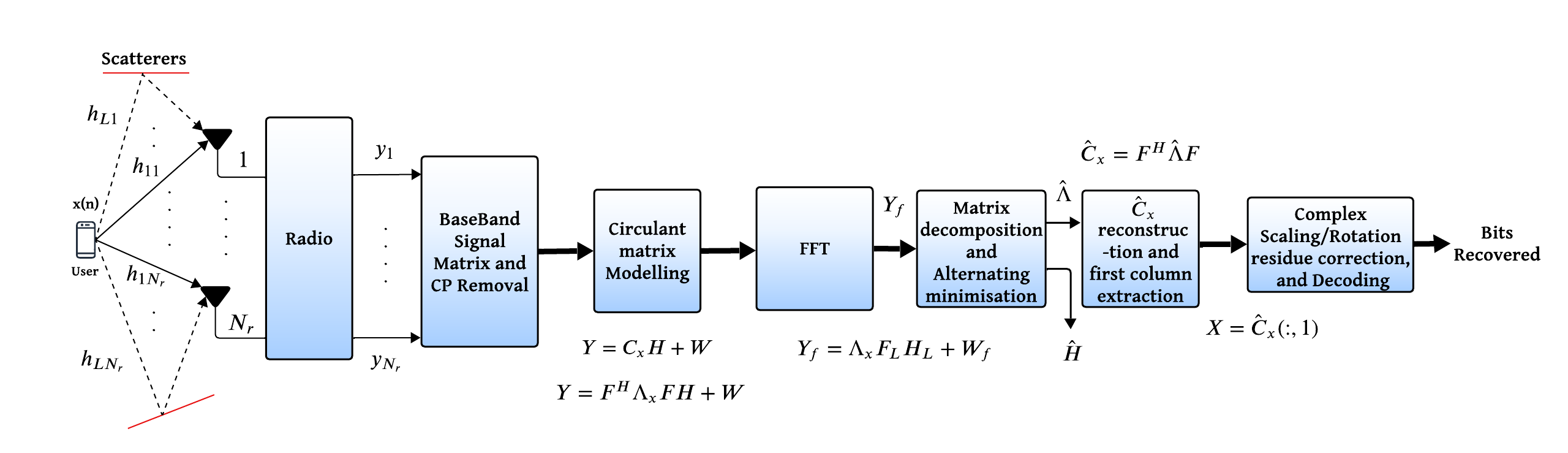}}
\caption{The Proposed receiver architecture for an SC-FDE System takes the baseband received signal from $N_r$ receive antennas after CP removal, converts it to a matrix and processes it through an FFT block. In the frequency domain, matrix decomposition and alternating minimization techniques are used to jointly estimate the channel and decode the transmitted data. The estimated signal component is used to reconstruct the circulant matrix, bringing the signal back to time domain. Finally, the resulting time-domain samples are corrected for scaling and rotation and the bits are recovered.}
\label{fig:block_diag}
\end{figure*}

Also, it should be noted that ${H}$ has only $L$ meaningful multi-paths at each receiver antenna in the channel. Thus the system can be modeled by considering $F_L$, the first $L$ columns of the DFT matrix ${F}$ and $H_L$, the first $L$ rows of the channel ${H}$.
\begin{equation} \label{eq:8}
    {Y = F^{H} \Lambda_x F_L H_L + W},
\end{equation}
Since the equalization is done in frequency domain for SC-FDE systems, we convert \eqref{eq:8} into frequency domain by pre-multiplying it with a $P\times P$ DFT matrix,
\begin{equation*}
    {FY = F F^{H} \Lambda_x F_L H_L + FW},
\end{equation*}
Since ${F}$ is unitary,
\begin{equation} \label{eq:final_Y_eqn}
    {Y_f = \Lambda_x F_L H_L + W_f}.
\end{equation}
where ${\Lambda_x}$ is a $P\times P$ diagonal matrix, ${F_L}$ is a $P\times L$ partial DFT matrix, ${H_L}$ is a $L\times N_r$ channel matrix, ${Y_f = FY}$ and ${W_f = FW}$ are the frequency domain received signal
and additive noise respectively. Figure \ref{fig:block_diag}, details the receiver architecture for the blind decoding algorithm using the proposed system model. The transmitted data $x[n]$ can be recovered once ${C_x}$ is known, which is known once the signal component ${\Lambda_x}$ is recovered. Throughout this paper, we assume that the channel has $L$ multi-paths/taps, perfect timing and frequency synchronization and perfect matched filtering at the receiver.

\section{Matrix Decomposition based Near-Pilotless Receiver} 
In conventional pilot-based systems employing frequency-domain equalization, pilot frames are transmitted to estimate the channel $H_L$, which is then used to decode subsequent frames. In contrast, this work assumes no pilot symbols except for a single time-domain pilot used for scaling correction. The objective is to recover the signal component $\Lambda_x$ without the aid of pilots, i.e., through blind decoding.
  
\subsection{Minimization Problem and Algorithm}

In this paper, we estimate $\Lambda_x$ and $H_L$ without any pilots by solving the following minimization problem.
\begin{equation} \label{eq:10}
    {(\hat{\Lambda},\hat{H}}) = \argmin_{\Lambda_x,H_L}{\lVert{Y_f - \Lambda_x F_L H_L\rVert}^{2}_{F}}.
\end{equation}
The problem in \eqref{eq:10} is non-convex jointly with $\Lambda_x$ and $H_L$, but convex individually. To solve this, we use alternating minimization, which alternately solves one variable at a time using the previous estimate of the other. This method is key to solving our blind decoding problem and is detailed in Algorithm \ref{alg:alt_min_fde}. We start with an initial guess $\hat{\Lambda}_0$ and the prior knowledge of the number of taps\footnote{
 Note that an estimate of number of taps $L$ can be obtained from preamble signaling. Preamble based correlation generates multiple peaks, each indicating a tap. As long as this estimate is $\geq L$, the algorithm performs well. We assume that we know the true values of $L$}. Subsequently, the method alternates between two minimization steps. In the first step, the channel is estimated by treating $\Lambda_x$ as known and by solving:

\begin{equation} \label{eq:min_eq_h}
    {\hat{H}} = \argmin_{H_L}{\lVert{Y_f - \Lambda_x F_L H_L\rVert}^{2}_{F}}.
\end{equation}
The solution to \eqref{eq:min_eq_h} is,  
\begin{equation} \label{eq:sol_eq}
    \hat{H} = ((A^H A + \mu I_L)^{-1} A^H Y_f).
\end{equation}
where, $A = \hat{\Lambda} F_L$ and $I_L$ is an $L\times L$ identity matrix. To ensure numerical stability, a regularization parameter $\mu$ is used. Based on extensive experimental findings, $\mu$ is selected to be within the range of $0<\mu<1$.
The second step is to utilize the estimated channel to solve,
 
\begin{equation} \label{eq:min_eq_x}
    {\hat{\Lambda}} = \argmin_{\Lambda_x}{\lVert{Y_f - \Lambda_x F_L H_L\rVert}^{2}_{F}}.
\end{equation}
The solution to \eqref{eq:min_eq_x} is given in Step-8 of Algorithm \ref{alg:alt_min_fde}. The proposed algorithm iteratively alternates between these two steps until the criterion specified in Step-11 of Algorithm \ref{alg:alt_min_fde} is met.
\renewcommand\footnoterule{}      
\begin{algorithm}[hbt!] 
\caption{Alternating Minimization (AM) for Blind Decoding using Matrix Decomposition}
\label{alg:alt_min_fde}
\begin{algorithmic}[1]
\State Input $\mathbf{Y_f}$, $L$ and define error tolerance ${\epsilon}$.
\State Initialize diagonal matrix ${\ {\hat{\Lambda}_0}}$ to be the top left singular column of $\ {Y_f}$ paired with correct column of DFT matrix and $n = 1$.
\Repeat
\State $A = \Lambda_{n-1} F_L $.
\State $\hat{H}_t = ((A^H A + \mu I_L)^{-1} A^H Y_f)$ .
\State $\hat{H}_n = F_L \hat{H}_t$.
\For{$p \gets 1 \text{ to } P$}
 \State $\hat{\Lambda}_n(p) = \frac{\sum_{r=1}^{N_r}{y(p,r){\hat{h}_n}^*(p,r)}}{\sum_{r=1}^{N_r}{\hat{h}_n(p,r){\hat{h}_n}^*(p,r)}}.$
\EndFor

 \State $n \gets n+1$
\Until{$\lVert  {Y_f} - {{{F}^H}}{{\hat{\Lambda} F_L}} \rVert_{2} / \lVert{Y_f}\rVert_{2} < \epsilon$}.\\
\Return $\hat{\Lambda}_n, \hat{H}_t$.
\end{algorithmic}
\end{algorithm}
 
 Upon termination, the AM method returns the estimated signal component ${\hat{\Lambda}_{n}}$ and ${\hat{H}_{t}}$. Using the obtained  ${\hat{\Lambda}_n}$ the corresponding circulant matrix is reconstructed again as described in \eqref{eq:5}, given by,
\begin{equation} \label{eq:circ_form}
    {\hat{C_x} = F^H \hat{\Lambda}_{n} F}.
\end{equation}
The IFFT operation after equalization in an SC-FDE system, is implicitly handled during the reconstruction of $\hat{C}_x$ as shown in \eqref{eq:circ_form}. Since $\hat{C_x}$ is a $P\times P$ circulant matrix, whose columns are circularly shifted versions of each other, we extract only the first column of $\hat{C_x}$ which serves as the estimate of the transmitted data $x[n]$. So,
\begin{equation} \label{eq:12}
    {X \gets \hat{C_x}(:,1)}.
\end{equation}
Plot $(A)$ in Fig. \ref{fig:rot_pilo_ca_qq_rr} shows the scatterplot corresponding to $X$, indicating scaling effect. This scaling is explained by Lemma 1 in \cite{10025791}, which states that the the output quantities $X$ and $\hat{H}_{n}$ of an alternative minimization algorithm are subject to scaling by a certain complex scalar ${\alpha}$, given by,
\begin{equation} \label{eq:13}
    {X_{de-rot} = \alpha{X}}, {H_{de-rot} = \frac{1}{\alpha}\hat{H}_{t}}.
\end{equation}
Estimating the complex scaling factor $\alpha$ is done using the pilot sample. 
\begin{equation}\label{eq:18}
    {{\alpha} = \frac{X[0]}{x[0]}},
\end{equation}
where, $x[0]$ is the transmitted pilot symbol and $X[0]$ is the first sample of the estimated data $X$. Once $\alpha$ is estimated, the recovered constellation is adjusted according to \eqref{eq:13}. However, a closer inspection of de-rotated constellation $X_{de-rot}$ as shown in plot $(B)$ of Fig. \ref{fig:rot_pilo_ca_qq_rr}, exhibits a noticeable tilt relative to the ideal QAM constellation structure. Section. \ref{sec:IV} explains the cause of this scaling/rotation residue and its mitigation.

\subsection{Choosing the Initial Point}
Given the non convex nature of our objective function, a good initial point is critical for the algorithm's performance, which can be either $\hat{\Lambda}$ or $\hat{H}$. However, the initial point chosen must not fall in the orthogonal subspace of the true quantities being estimated, as the algorithm will not converge to true quantities $\Lambda_x$ and $H_L$. This is because no transformation (scaling or rotation) can project a vector onto a subspace it is orthogonal to \cite{orthogonal_convergence}. To ensure the chosen initial point lies in the same sub-space as the true quantity, we construct the initial value $\hat{\Lambda}_0$, using the top left singular vector of $Y_f$. The singular value decomposition (SVD) of $Y_f$ is given as
   \begin{equation}\label{eq:svd_decomp}
     {Y_f = U_{P \times P} \Sigma_{P\times Nr} V^H_{N_r\times N_r}},
    \end{equation}
Equating \eqref{eq:final_Y_eqn} and \eqref{eq:svd_decomp}, we have
    \begin{equation}\label{eq:svd}
        {Y_f = U\Sigma V^H = \Lambda_x F_L H_L }.
    \end{equation}
From \eqref{eq:svd}, the column space corresponding to $\Lambda_x$ and $U$ are same, making the top left singular vector a good choice for constructing the initial value $\hat{\Lambda}_0$. The top left singular vector paired with the correct column of DFT matrix \cite{my_init_pt_paper} is chosen as an initial point. The analysis of Algorithm \ref{alg:alt_min_fde}'s sensitivity to the chosen initial point is discussed in detail in \cite{my_init_pt_paper}.

\section{Scaling and Rotational Residue Correction}\label{sec:IV}
To compensate the rotation, a single pilot \eqref{eq:13} is used. However, the noise in the pilot may cause residual rotation. Eliminating this residue improves the performance in terms of error rates. We propose two methods for correcting scaling and rotation residue respectively.
\subsection{Scaling Correction using Centroids Adjustment (CA) method}

The proposed method bypasses ${X_{de-rot}}$ and instead uses the returned signal ${X}$ from \eqref{eq:12}. 
Aligning ${X}$ with standard QAM points is difficult due to unpredictable rotation. A simpler approach is to select a maximum value from $X$ and scale it relative to a known standard QAM point to obtain $X_{mid}$.  However, using only the maximum value rather than the centroid of the data cluster leaves some residual rotation. To overcome this, demodulate $X_{mid}$ and identify indices that match the demodulated index of the previously selected maximum value.

\begin{figure}[ht!]
    \centering
    \includegraphics[width = 0.48\textwidth]{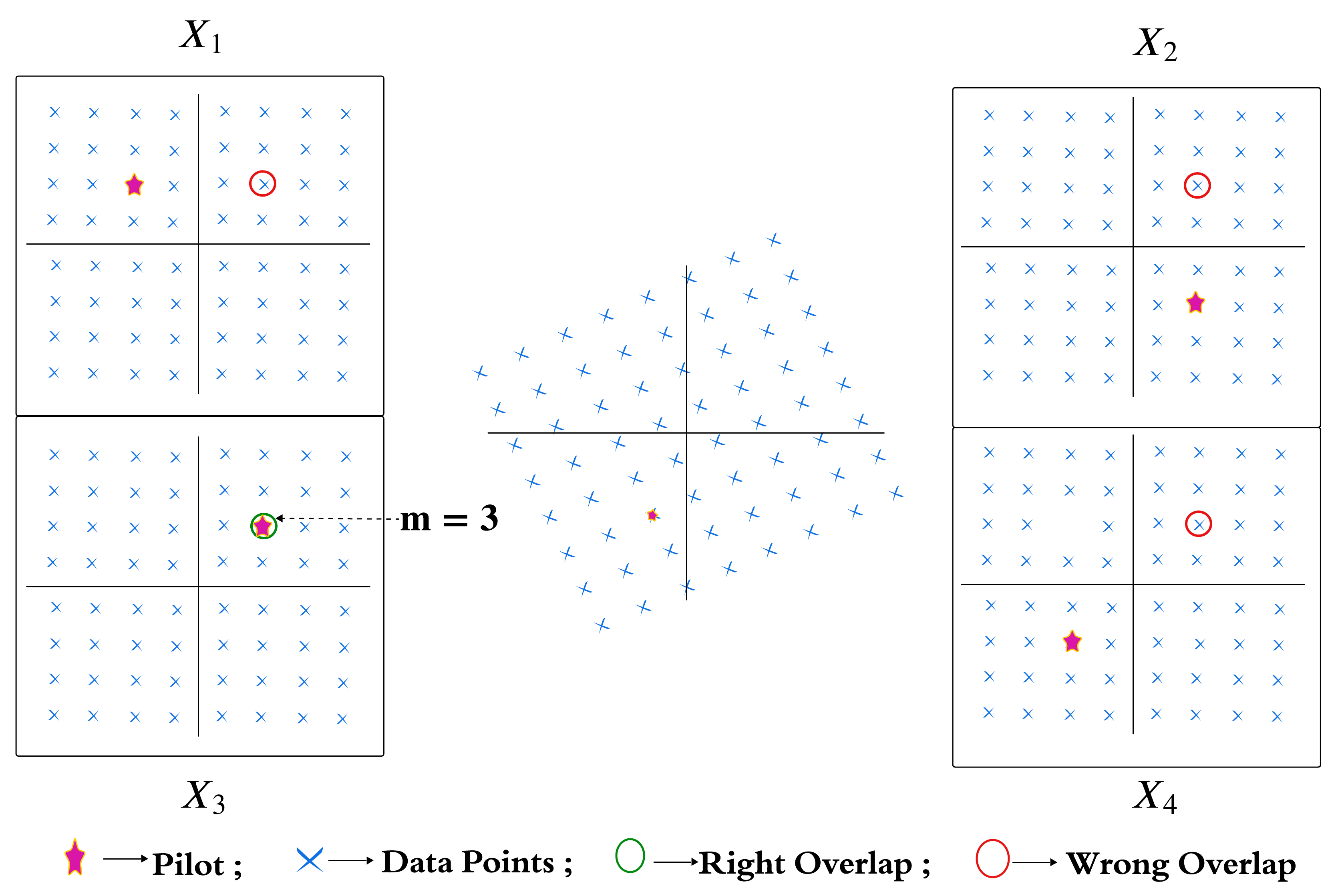}
    \caption{Centroids adjustment based scaling correction: The figure shows estimation of $m$ which tells us the right constellation among four output constellations by making use of the single pilot}
    \label{fig:kc_rr}
\end{figure}

By averaging all the data points that have matching indices, we compute the centroid and correct it with the maximum centroid value of a true QAM constellation in each quadrant. This simplifies the problem to determining four possible angles, each a multiple of $\frac{\pi}{2}$ rotation due to the QAM's symmetrical square structure. We obtain $m\frac{\pi}{2}$ rotations, where $m \in [0,1,2,3]$ of ${X_{q}}$, with only one value of $m$ representing the correct alignment as shown in Fig. \ref{fig:kc_rr}. The correct value of $m$ is estimated from the pilot symbol $x[0]$ of the transmitted sequence. The centroid adjustment is summarized in Algorithm \ref{alg:centroids adjustment_algo}, and the resulting scatterplot of $X_{true}$ is depicted in plot $(C)$ of Fig. \ref{fig:rot_pilo_ca_qq_rr}.

\begin{algorithm}[ht] 
\caption{Centroids Adjustment (CA) Algorithm for Scaling Correction}
\label{alg:centroids adjustment_algo}
\begin{algorithmic}[1]
\State Input $\mathbf{X}$.

\State Define $\mathcal{C}_M = \{ c \in \mathbb{C} \mid c \in \text{centroids of M-QAM} \}$.

\State Define $\mathcal{C}_{M,q}=\{c_q \in \mathbb{C} \mid c_q \in \text{centroids of $\mathcal{C}_M$ per $q$}$\}. where, $q={1,2,3,4}$ denotes quadrants of a 2D plane.


\State Compute $\alpha_{mid} = \frac{\max(X)}{\max(\mathcal{C}_{M,1})}$ and corresponding $X_{mid} = X\odot\frac{1}{\alpha_{mid}}$.
\State Define $D \gets$ M-QAM demod $(X_{mid}$) and $D_{max} \gets$ M-QAM demod ($ \max (X_{mid}$)).
\State ${J = \index1(D = D_{max})}$ and estimate the centroid as ${\mathcal{C}_{X_{mid}} = \frac{1}{\size(J)}\sum{X_{mid}(J)}}$.
\For{$q \gets 1 \text{ to } 4$}
 \State $b_q = \max(\mathcal{C}_{M,q}).$
 \State $\alpha_{ca,q} = \frac{\mathcal{C}_{X_{mid}}}{b_q}.$
 \State ${X_q} = X\odot{\frac{1}{\alpha_{ca,q}}}.$
\EndFor 
\State $m = \argmin_{q}{\lVert{X_q[0]} - x[0]}\rVert^{2}_{2}.$\\
\Return  ${X_{true} = X_m}$.
\end{algorithmic}
\end{algorithm}


\subsection{Rotational Residue (RR) correction using QAM to QPSK (QQ) method} 
\begin{figure}[ht]
    \centering
    \includegraphics[width = 0.5\textwidth]{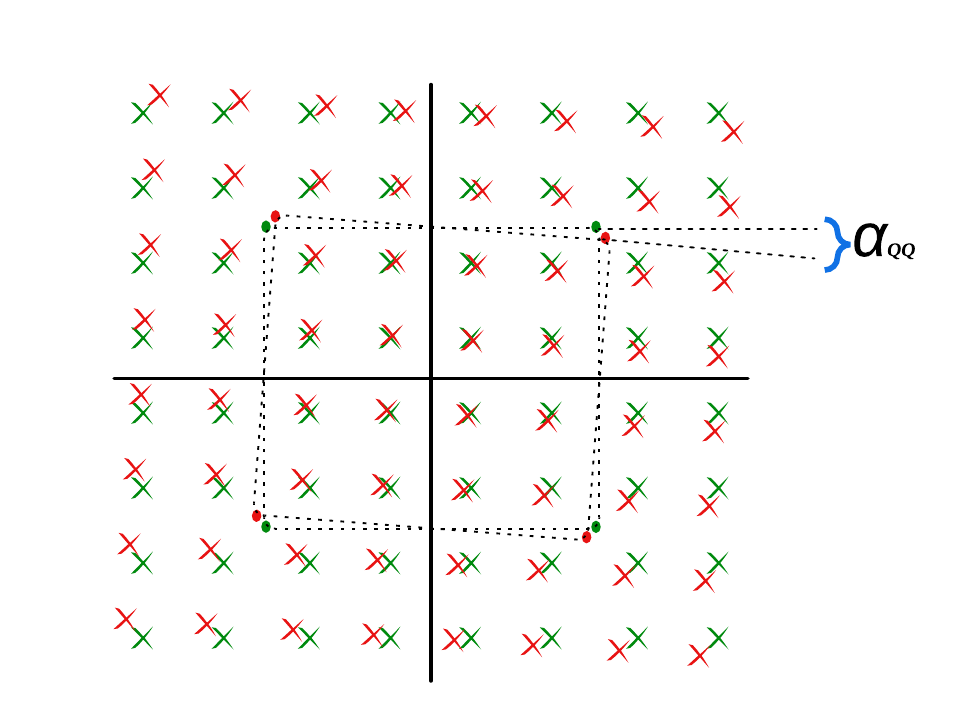}
    \caption{QAM to QPSK based RR correction: The example depicts a 64 QAM constellation with RR introduced. The $\alpha_{QQ}$ corresponds to the offset between the actual location of QPSK points (green points) and the residue point (red points)}
    \label{fig:qq_rr}
\end{figure}

\begin{figure}[ht]
   \centering
    \includegraphics[width = 0.5\textwidth]{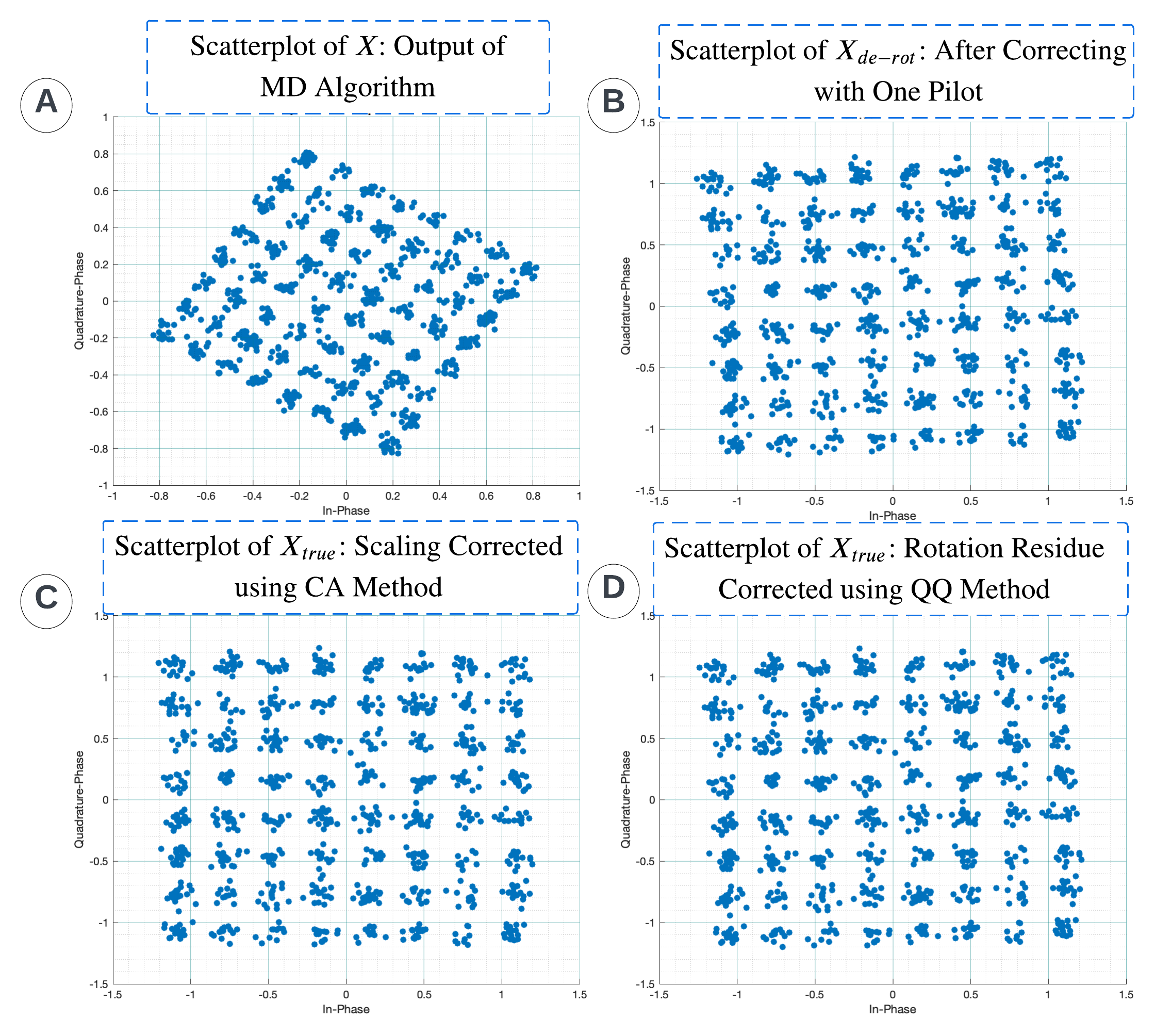}
    \caption{At 7 dB SNR, $P=1024$ , $N_r = 64$, $L=9$ for 64 QAM - Scatter-plots of : \textbf{(A)}. Estimated ${X}$ from alternating minimization algorithm after IFFT block, \textbf{(B)}. Output ${X_{de-rot}}$ after correcting ${X}$ with one pilot, \textbf{(C)}. Signal ${X_{true}}$ obtained by correcting scaling using Centroids Adjustment (CA) method, \textbf{(D)}. Signal ${X_{true}}$ obtained by correcting the rotational residue using QAM to QPSK (QQ) method.}
   \label{fig:rot_pilo_ca_qq_rr}
\end{figure}

This subsection describes a simple yet effective rotational residue method called QAM to QPSK as shown in Algorithm \ref{alg:QQ_algo}. The method takes a pilot corrected constellation, ${X}_{de-rot}$ from \eqref{eq:13}, as input. This method outlined in Fig. \ref{fig:qq_rr}, maps the mean of all samples in quadrant $q$ of ${X}_{de-rot}$ to a known $q^{th}$-centroid of QPSK constellation. Each point in the QPSK constellation lies on a $\frac{\pi}{4}+(q-1)\frac{\pi}{2}$ angled line, where $q = [1,2,3,4]$. The offsets between the known QPSK positions and the mean values of ${X_{de-rot}}$ per quadrant are calculated and averaged. This average value is used to correct the rotational residue and obtain ${X_{true}}$, whose scatterplot is shown in plot $(D)$ of Fig. \ref{fig:rot_pilo_ca_qq_rr}.

\begin{algorithm}[ht!] 
\caption{QAM to QPSK (QQ) Algorithm for Rotational Residue Correction}
\label{alg:QQ_algo}
\begin{algorithmic}[1]
\State Input $\mathbf{X_{de-rot}}$.

\State Define $\mathcal{C}_{4,q}=\{c_q \in \mathbb{C} \mid c_q \in \text{$q$-centroid of QPSK}$\}.

\State Define $\mathcal{C}_{X,q} \gets \text{mean of all samples of}$ $X_{de-rot}$ of quadrant $q$. where $q={1,2,3,4}$ indicates quadrants.

\For{$q \gets 1 \text{ to } 4$}
\State $\eta_{q} = \frac{\mathcal{C}_{4,q}}{\mathcal{C}_{X,q}}$.
\EndFor 
\State $\alpha_{QQ} = \frac{1}{4} \sum_{q} \eta_{q}$.\\
\Return ${X_{true} = X_{de-rot}\odot\frac{1}{\alpha_{QQ}}}$.
\end{algorithmic}
\end{algorithm} 

\section{Performance Evaluation and Results}
The system performance is evaluated for an uncoded single carrier system using the simulation parameters\footnote{The TDLA-30 channel model considered comprises of $L$ multi-paths/taps, with the first tap having the highest power and the subsequent taps having progressively lower powers. All multi-path components are assumed to have integer delay values} shown in Table. I. 
\begin{table}[h]\label{tab:sim_params1}
    \centering
    \renewcommand{\arraystretch}{1.1} 
    \begin{tabular}{|c|c|}
        \hline
        \textbf{Parameter} & \textbf{Value} \\
        \hline
        Channel & TDLA-30 \\
        Modulation Order ($M$) & 64-QAM \\
        Tx sequence length ($P$) & 1024 \\
        Number of receiver antennas ($N_r$) & 64 \\
        Number of multi-paths/taps ($L$) & 9 \\
        AM Iterations to reach $\epsilon$ & 20 \\
        Regularization factor ($\mu$) & 0.5 \\
        \hline
    \end{tabular}
    
    \vspace{0.5em} 
    {\small {Table I:} Simulation Parameters} 
\end{table}

For comparison, an MRC-based OFDM system is used as the baseline. For OFDM, the channel estimation is performed conventionally using $10\%$ equidistant placed pilots in the transmitted sub-carriers, with FFT based interpolation. 
\begin{figure}[hbt!]
    \centering
    \includegraphics[scale=0.44]{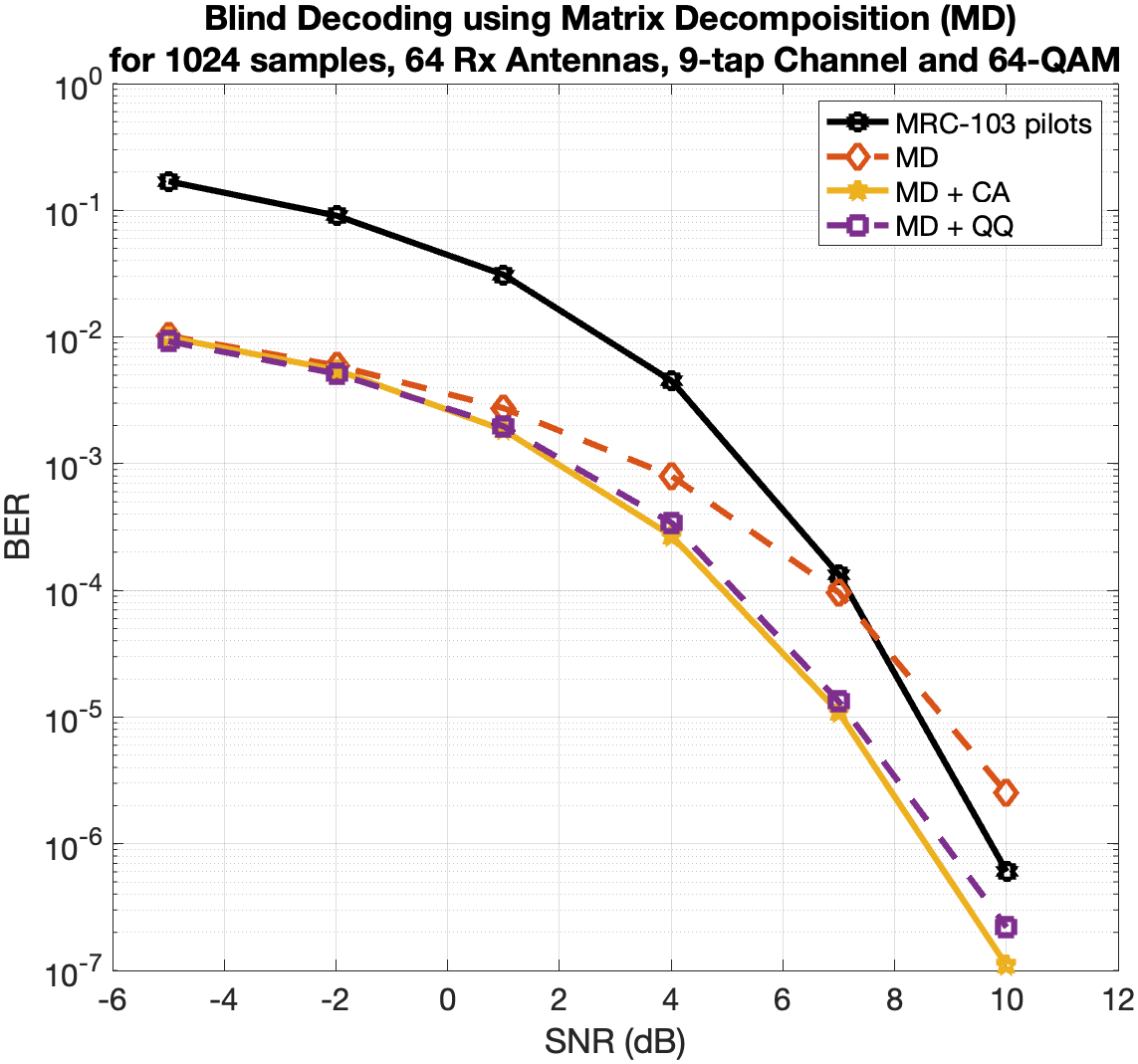}
    \caption{Comparison of uncoded BER of MRC using $10\%$ pilots and the proposed blind matrix decomposition algorithm.}
    \label{fig:ber_64qam}
\end{figure}
In  Fig. \ref{fig:ber_64qam}, the bit error rate (BER) is plotted against SNR (dB). We observe from Fig. \ref{fig:ber_64qam} that at lower SNRs, the proposed algorithm outperforms MRC with $10\%$ pilots. However, as SNR increases, the algorithm’s performance deteriorates due to the significant impact of estimating the complex scalar $\alpha$ using noisy pilots. One way to improve the performance is to increase the number of pilots and average the noise. Since the idea is to reduce the overhead, we use either the scaling correction using CA or rotational residue correction using QQ methods, which leads to an improved performance in terms of bit error rate as shown in Fig. \ref{fig:ber_64qam}.

\begin{figure}[ht!]
    \centering
    \includegraphics[scale=0.44]{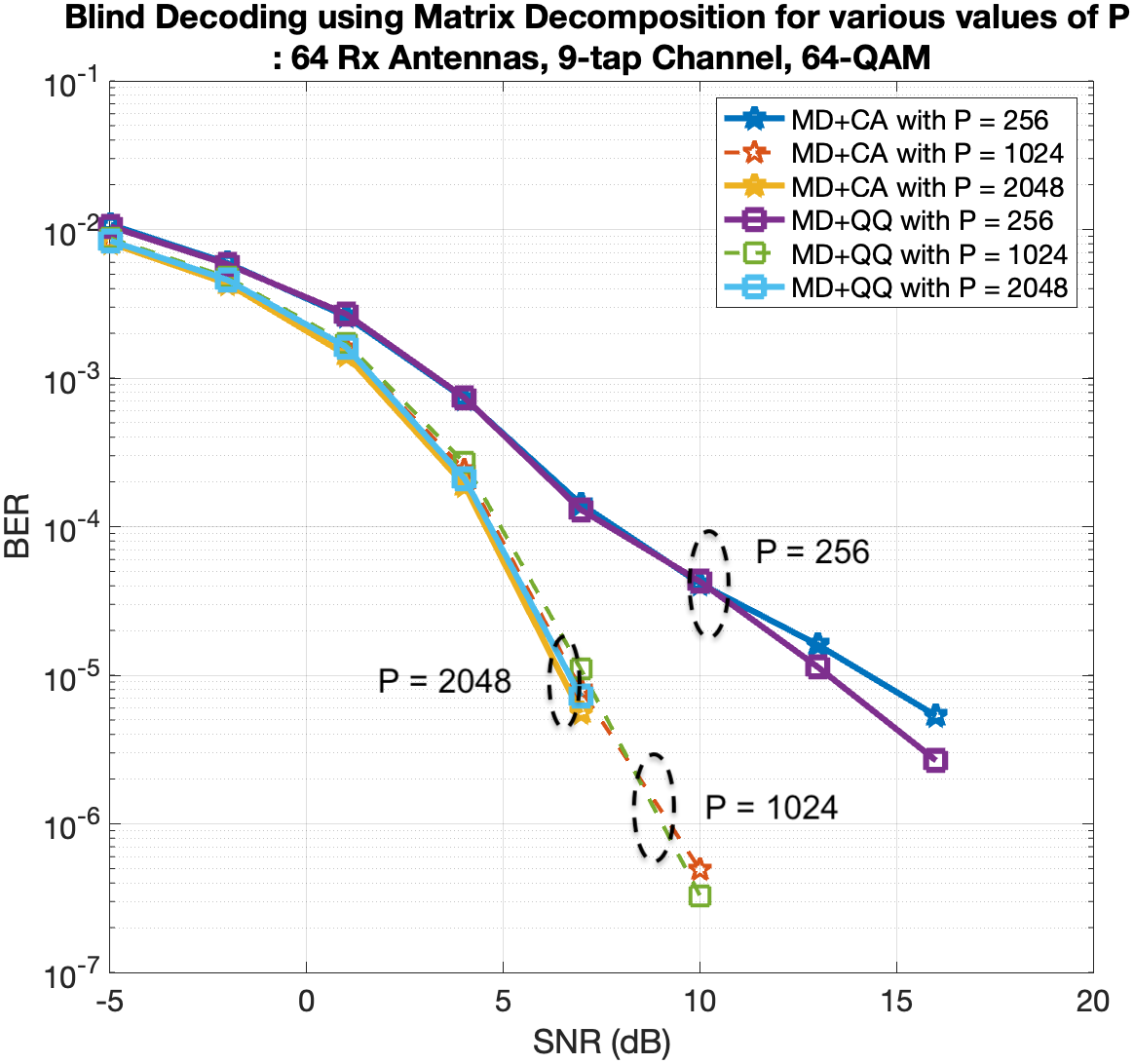}
    \caption{Comparison of blind decoding algorithm with various values of $P$ using CA based scaling correction and QQ based residual correction.}
    \label{fig:P_comps_plot}
\end{figure}

\begin{figure}[ht!]
    \centering
    \includegraphics[scale=0.45]{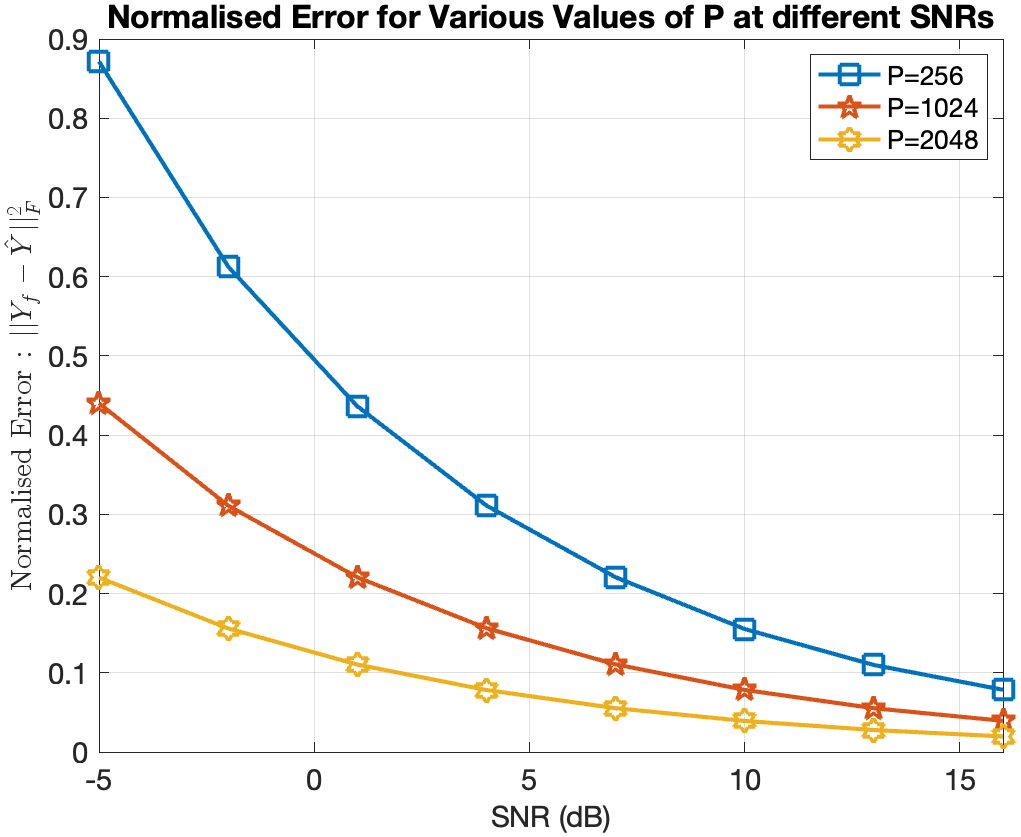}
    \caption{Normalized error for various values of $P$ using parameters shown in Table. I. Here, the reconstructed signal $\hat{Y} = F^H \hat{\Lambda}_n F_L \hat{H}_t$} 
    \label{fig:error_plot}
\end{figure}

It is evident from Fig. \ref{fig:P_comps_plot} that as $P$ increases\footnote{The value of $P$ is chosen to be a power of 2.}, the algorithm's accuracy improves in terms of bit error rate. Both the scaling and residual correction methods demonstrate comparable performance with improved BER as $P$ increases. This is supported by the normalized error plot in Fig. \ref{fig:error_plot} showing decay for various values of $P$. The error between the true signal ($Y_f$) and the reconstructed signal ($\hat{Y}$) improves with increasing $P$, attributable to the gains provided by MRC during equalization.

\subsection*{Algorithm Complexity Analysis}
The computational complexity of AM algorithm can be expressed as a function of $P$, $L$ and $N_r$ with a consideration that $P \gg N_r \gg L $. The complexity of matrix pseudo inverse involving a diagonal matrix is detailed and derived in \cite{10025791}. Most of the complexity comes from Step 5 of Algorithm \ref{alg:alt_min_fde} which is of $\mathcal{O}(PLN_r)$, indicating a direct proportionality to the sequence length $P$. The approximate complexity per iteration of the AM algorithm is $\mathcal{O}(PLN_r)$, which is similar to the complexity $\mathcal{O}(P^3)$ of the SVD update rule in PCA compression \cite{7511265}, but more computationally intensive because it requires multiple iterations. The number of samples in a sequence for one instance of the algorithm depends on the coherence time of the channel. Therefore, there is a trade-off between the improved accuracy and increased complexity resulting from longer sequence lengths. This method of matrix decomposition is performed for every $P$ length sequence transmission. If coherence time is large enough to maintain the channel estimates, the same estimates can be used for successive transmissions, reducing channel estimation and data decoding computations.

\section{Conclusion and Future Work} 
In this work, we propose a near-pilotless decoding algorithm for single carrier systems. The proposed iterative performs blind decoding using matrix decomposition with a single pilot, eliminating the need for prior channel or data knowledge. This approach reduces the pilot overhead while simultaneously providing estimates for both the signal and the channel. Matrix decomposition with rotation/scaling corrections does not experience a reduction in performance compared to MRC based equalization. The proposed method can be extended to multi-user systems,  with future work focusing on channel scenarios with two or more dominant tap powers.

\section*{Acknowledgment}
This work was funded by the Ministry of Electronics and Information Technology (MeitY), Government of India, through the project, “Next Generation Wireless Research and Standardization on 5G and Beyond”, and by ANSYS Software Pvt. Ltd. through their Doctoral Fellowship Award Program. 



\bibliographystyle{IEEEtran}

\bibliography{this_pap} 

@ARTICLE{995852,
  author={Falconer, D. and Ariyavisitakul, S.L. and Benyamin-Seeyar, A. and Eidson, B.},
  journal={IEEE Communications Magazine}, 
  title="{F}requency {D}omain {E}qualization for {S}ingle-{C}arrier {B}roadband {W}ireless {S}ystems", 
  year={2002},
  volume={40},
  number={4},
  pages={58-66},
  doi={10.1109/35.995852}}

@INPROCEEDINGS{7511265,
  author={Choi, Jinseok and Evans, Brian L. and Gatherer, Alan},
  booktitle={2016 IEEE International Conference on Communications (ICC)}, 
  title="{Space-time Fronthaul Compression of Complex Baseband Uplink LTE Signals}", 
  year={2016},
  volume={},
  number={},
  pages={1-6},
  doi={10.1109/ICC.2016.7511265}}

@INPROCEEDINGS{tomasi2016pilot,
  author={Tomasi, Beatrice and Guillaud, Maxime},
  booktitle={2015 49th Asilomar Conference on Signals, Systems and Computers}, 
  title="{Pilot Length Optimization for Spatially Correlated Multi-User MIMO Channel Estimation}", 
  year={2015},
  volume={},
  number={},
  pages={1237-1241},
  doi={10.1109/ACSSC.2015.7421339}}

@article{yuan2024integratednearfieldsensing,
  title="{Integrated Near Field Sensing and Communications Using Unitary Approximate Message Passing Based Matrix Factorization}",
  author={Yuan, Zhengdao and Guo, Qinghua and Eldar, Yonina C and Li, Yonghui},
  journal={arXiv preprint arXiv:2406.07272},
  year={2024}
}

@book{golub13,
  author = {Golub, Gene H. and van Loan, Charles F.},
  edition = {Fourth},
  isbn = {1421407949 9781421407944},
  publisher = {JHU Press},
  refid = {824733531},
  title = {Matrix Computations},
    year = 2013
}

@ARTICLE{10025791,
  author={Aswathylakshmi, P. and Ganti, Radha Krishna},
  journal={IEEE Open Journal of the Communications Society}, 
  title="{Fronthaul Compression for Uplink Massive MIMO Using Matrix Decomposition}", 
  year={2023},
  volume={4},
  number={},
  pages={518-533},
  doi={10.1109/OJCOMS.2023.3238772}}

@inproceedings{Acolatse2010DesignOS,
  title="Design of {S}ingle-{C}arrier {F}requency {D}omain {E}qualization ({SC-FDE}) with {T}ransmit {D}iversity for {W}ireless and {O}ptical {C}ommunications",
  author={Kodzovi Acolatse},
  year={2010},
  note="{A}ccessed: 23 Sep, 2024",
}

@ARTICLE{1369274,
  author={Yonghong Zeng and Tung Sang Ng},
  journal={IEEE Signal Processing Letters}, 
  title="{Pilot Cyclic Prefixed Single Carrier Communication: Channel Estimation and Equalization}", 
  year={2005},
  volume={12},
  number={1},
  pages={56-59},
  doi={10.1109/LSP.2004.839698}}

@ARTICLE{935746, 
  author={Deneire, L. and Gyselinckx, B. and Engels, M.},
  journal={IEEE Communications Letters}, 
  title="{Training Sequence versus Cyclic Prefix-a New Look on Single Carrier Communication}", 
  year={2001},
  volume={5},
  number={7},
  pages={292-294},
  doi={10.1109/4234.935746}}

@INPROCEEDINGS{7485388,
  author={Cai, Jing-Jing and Su, Wei and Zhang, Sheng-nan and Chen, Ke-Yu and Wang, De-Qing},
  booktitle={OCEANS 2016 - Shanghai}, 
  title="{A Semi-Blind Joint Channel Estimation and Equalization Single Carrier Coherent Underwater Acoustic Communication Receiver}", 
  year={2016},
  volume={},
  number={},
  pages={1-6},
  doi={10.1109/OCEANSAP.2016.7485388}}

@INPROCEEDINGS{4773788,
  author={Chen, Yi-Sheng},
  booktitle={2008 14th Asia-Pacific Conference on Communications}, 
  title="{Blind Channel Estimation for Single Carrier Block Transmission Systems with Frequency Domain Equalization}", 
  year={2008},
  volume={},
  number={},
  pages={1-5}}

@ARTICLE{zhang2017blindsignaldetectionmassive,
  author={Zhang, Jianwen and Yuan, Xiaojun and Zhang, Ying-Jun Angela},
  journal={IEEE Transactions on Communications}, 
  title="{Blind Signal Detection in Massive MIMO: Exploiting the Channel Sparsity}", 
  year={2018},
  volume={66},
  number={2},
  pages={700-712},
  doi={10.1109/TCOMM.2017.2761384}}

@inproceedings{Aswathylakshmi_2023,
   title="{Pilotless Uplink for Massive MIMO Systems}",
   DOI={10.1109/globecom54140.2023.10437723},
   booktitle={GLOBECOM 2023 - IEEE Global Communications Conference},
   publisher={IEEE},
   author={Aswathylakshmi, P. and Ganti, Radha Krishna},
   year={2023},
   month=dec, pages={4205–4210} }

@article{orthogonal_convergence,
author = {Beck, Amir},
title = "{On the Convergence of Alternating Minimization for Convex Programming with Applications to Iteratively Reweighted Least Squares and Decomposition Schemes}",
journal = {SIAM Journal on Optimization},
volume = {25},
number = {1},
pages = {185-209},
year = {2015},
doi = {10.1137/13094829X},
}

@INPROCEEDINGS{my_init_pt_paper,
  author={Praneeth, K. Sai and Aswathylakshmi, P. and Ganti, Radha Krishna},
  booktitle={2025 National Conference on Communications (NCC)}, 
  title="{An Initial Point for Blind Decoding of OFDM Using Matrix Decomposition}", 
  year={2025},
  volume={},
  number={},
  pages={1-6}}

\end{document}